\documentstyle{article}
\begin{document}
\LARGE
\begin{center}
\bf  Open and Closed Creations of Black Hole Pair 
\vspace*{0.6in}
\normalsize \large \rm 

Wu Zhong Chao

Dept. of Physics

Beijing Normal University

Beijing 100875, China

(Oct. 24, 1998)

\vspace*{0.4in}
\large
\bf
Abstract
\end{center}
\vspace*{.1in}
\rm
\normalsize
\vspace*{0.1in}

In the absence of a general no-boundary proposal for open
creation, the constrained instanton approach is used in treating
both the open and closed pair creations of black holes. A 
constrained instanton is considered as the seed for the
quantum pair creation of black holes in the Kerr-Newman-(anti-)de
Sitter family. At the $WKB$ level, for the chargeless and
nonrotating case, the creation probability is the exponential of
the (minus) entropy of the universe. Also for the other cases
(charged, rotating, or both), the creation probability is the
exponential of (minus) one quarter of the sum of the inner and
outer black hole horizon areas. The case of the Kerr-Newman
family is also solved as a limiting case of that for the 
Kerr-Newman-anti-de Sitter family. The study of the open creation
of a black hole pair can be considered as a prototype of the
constrained instanton method to quantum gravity for an open
universe, without appealing to the background subtraction
approach.

\vspace*{0.3in}

PACS number(s): 98.80.Hw, 98.80.Bp, 04.60.Kz, 04.70.Dy

Keywords: quantum cosmology, constrained gravitational instanton,
black hole creation

\vspace*{0.3in}

e-mail: wu@axp3g9.icra.it

\pagebreak

\vspace*{0.3in}

\large \bf I. Introduction
\vspace*{0.15in}

\rm 

\normalsize

In the No-Boundary Universe, the wave function of a closed
universe is  defined as
a path integral over all compact 4-metrics with matter fields
[1]. The dominant contribution to
the path integral is from the stationary action solution. At
the $WKB$ level, the wave function
can be written as
\begin{equation}
\Psi \approx e^{- I},
\end{equation}
where $I= I_r + iI_i$  is the complex action of the solution.

The Euclidean action  is
\begin{equation}
I = - \frac{1}{16 \pi} \int_M (R - 2 \Lambda + L_m) - \frac{1}{8
\pi}
\oint_{\partial M } K,
\end{equation}
where $R$ is the scalar curvature of the spacetime $M$, $K$
is the trace of the second form of the boundary $\partial M$,
$\Lambda$ is the cosmological constant, and
$L_m$ is the Lagrangian of the matter content.

The imaginary part $I_i$ and real part $I_r$ of the
action represent the Lorentzian
and Euclidean evolutions in real time and imaginary time,
respectively. When their orbits are intertwined they are mutually
perpendicular in the configuration space with the supermetric.
The probability of a
Lorentzian orbit remains constant during the evolution. One can
identify the probability, not only
as the probability of the universe created, but also as
the probabilities for other Lorentzian
universes obtained through an analytic continuation from it [2].

An instanton is defined as a stationary action orbit and
satisfies the Einstein equation everywhere, and is
the seed for the creation of the universe. However, very few
regular instantons exist. The
framework of the No-Boundary Universe is much wider than that of
the instanton theory. Therefore, in order not to exclude many
interesting phenomena from
the study, one has to appeal
to the concept of constrained instantons [3]. Constrained
instantons are the orbits
with an action which is stationary
under some restriction. The restriction can be imposed
on a spacelike 3-surface of the created Lorentzian
universe. This restriction is that the 3-metric and matter
content are given at the 3-surface. The relative creation
probability from the
instanton is the exponential of the negative of the
real part of the instanton action.

The usual prescription for finding a constrained instanton is to
obtain a complex solution to the Einstein equation and other
field equations in the complex domain of spacetime coordinates.
If there is no singularity in a compact section of the solution,
then the section is considered as an instanton. If there exist
singularities in the section, then the action of the section is
not
stationary. The action may only be stationary with respect to the
variations under some restrictions mentioned above. If this is
the case, then the section is a constrained gravitational
instanton. To find the constrained instanton, one has to closely
investigate the singularities. The stationary action condition is
crucial to the validation of the $WKB$ approximation. We are
going to work at the $WKB$ level for the problem of quantum
creation of a black hole pair.

A main unresolved problem in quantum cosmology is to
generalize the no-boundary proposal for an open universe. While a
general prescription  is not available, one can still use 
analytic continuation to obtain the $WKB$ approximation to the
wave function for open universes with some kind of
symmetry.

The most symmetric space is the $S^4$ space with $O(5)$ symmetry,
or the four-sphere,
\begin{equation}
ds^2 = d\tau^2 + \frac{3}{\Lambda} \cos^2 \left (
\sqrt{\frac{\Lambda}{3}} \tau \right ) (d \chi^2
+ \sin^2 \chi (d \theta^2 + \sin^2 \theta d \phi^2)) ,
\end{equation}
where $\Lambda$ is a positive cosmological constant. One can
obtain the
de Sitter space with 3-spheres as the spatial sections of
constant $t$ by the substitution $\tau = it$. The de Sitter space
with 3-hyperboloids as the spatial sections of constant $t$ is
obtained by the substitutions $ \tau =\sqrt{\frac{3}{\Lambda}}
\frac{\pi}{2} + it$ and $
\chi = i\rho$.

One can also obtain the anti-de
Sitter space by the substitution $ \chi = i\rho$. The signature
of  anti-de Sitter space is $(+,
-, -, -)$. This signature associated
with the anti-de Sitter space is reasonable, since the relative
sign of the cosmological constant is
implicitly changed by the analytic continuation. If one prefers
the usual signature of the anti-de
Sitter space, then he could start from the four-sphere with the
signature of $(-, -, -, -)$, instead of (3) [4].

One can reduce the $O(5)$ symmetry to make the model more
realistic.
This is the $FLRW$ space with $O(4)$ symmetry,
\begin{equation}
ds^2 = d\tau^2 + a^2(\tau )(d \chi^2 + \sin^2 \chi (d \theta^2 +
\sin^2 \theta d \phi^2)),
\end{equation}
where $a(\tau )$ is the length scale of the homogeneous
3-surfaces and $a(0) = 0$.   One can apply the combined analytic
continuations $t = -i\tau$ and $\chi = i\rho$ to obtain an open
$FLRW$ universe [2].

The study of the perturbation modes around this background,
strictly following the no-boundary philosophy, is waiting for a
general proposal for
the quantum state of an open universe. If one includes 
``realistic'' matter fields in the model, then the instanton is
not regular. However, the singular instanton can be interpreted
as a constrained instanton [5].

In this paper, we try to reduce the symmetry further, that is to
investigate vacuum models with $O(3)$, or
spherical symmetry. This will lead to a quantum pair creations
of black holes in the de Sitter (anti-de Sitter) space
background. We shall consider the vacuum model with a positive
cosmological constant first. There exist only two regular
instantons, they are the $S^4$ and $S^2 \times S^2$ spaces. These
are the origins of the de Sitter space and the Nariai
space [6]. The Nariai space is interpreted as a pair of black
holes with maximal mass $m_c = \Lambda^{-1/2}/3$. For the case of
sub-maximal black hole pair creation, one has to use a real
constrained
gravitational instanton as the seed for the creation. If the
cosmological constant is negative, one has to use a complex
constrained instanton as the seed for a pair creation of
Schwarzschild-anti-de Sitter black holes.  

This argument is generalized to the pair creations of all other
members
of the Kerr-Newman black hole family. 
For the pair creation of black holes in the de Sitter space
background, the real constrained instantons becomes
the seeds, while for the pair creation of black holes in the
anti-de Sitter space
background, one has to use the complex constrained instantons as
the seeds. We
shall investigate the pair creation of the Schwarzschild,
Reissner-Nordstr$\rm\ddot{o}$m, Kerr and Newman black holes in
Sects. II, III,
IV and V, respectively. For the rotating black hole case, the
symmetry has been reduced to $O(2)$. All these instantons have
an extra $U(1)$ isometry that agrees with the time invariance.

The pair creation probability of Schwarzschild black holes in the
closed de Sitter space background is the exponential of the
entropy of the universe, while in the open anti-de Sitter
space background the probability is the exponential of the
negative of the
entropy. This is reasonable, since it implies that the more
massive the black holes are, the more unlikely they are to be
created.

For the pair creations of other members in the Kerr-Newman family
in the closed (open) backgrounds, the
creation probability is the exponential of (minus) one quarter of
the sum of the outer and inner black hole horizon areas. The de
Sitter (anti-de Sitter) spacetime without a black hole is the
most probable evolution comparing with that with a pair of black
holes.

Sect. VI will discuss gravitational thermodynamics using
constrained instantons. The independence of the action from
the identification time period implies that the action is equal
to the negative of the entropy. However, one has to clarify the
physical meaning of the entropy. Sect.VII is a discussion. It is
argued that to respect the principle of general covariance, one
has to work in the domain of complex coordinates in quantum
gravity. This justifies the quantum creation of a black hole pair
from our complex constrained instanton.

\vspace*{0.3in}

\large \bf II. Schwarzschild
\vspace*{0.15in}

\rm 

\normalsize

The  solution to the Einstein equation is written
\begin{equation}
ds^2 = \Delta d\tau^2 + \Delta^{-1} dr^2 + r^2( d\theta^2 +
\sin^2 \theta d\phi^2 ),
\end{equation}
\begin{equation}
\Delta = 1 -\frac{2m}{r} -  \frac{\Lambda r^2}{3},
\end{equation}
where $m$ is an integral constant.  One can make a factorization 
\begin{equation}
\Delta = - \frac{\Lambda}{3r}
(r - r_0)(r - r_2)(r - r_3).
\end{equation}
One can define
\begin{equation}
\alpha = \frac{1}{3} \mbox{arccos}(3m\Lambda^{1/2}),
\end{equation}
then one has
\begin{equation}
r_2 = 2 \sqrt{\frac{1}{\Lambda}} \cos \left ( \alpha +
\frac{\pi}{3} \right ), \;\;r_3 = 2 \sqrt{\frac{1}{\Lambda}} \cos
\left ( \alpha - \frac{\pi}{3} \right ),
\end{equation}
\begin{equation}
r_0 = -2 \sqrt{\frac{1}{\Lambda}} \cos \alpha .
\end{equation}

The  surface gravity $\kappa_i$ of $r_i$ is [7] 
\begin{equation}
\kappa_i = \frac{\Lambda}{6r_i}\prod_{j = 0,2,3, \;\; (j
\neq i)}(r_i - r_j).
\end{equation}

If $\Lambda$ is positive and $0 \le m \le m_c $,
then $r_2$ and $r_3$ are
real. There exist regular instantons $S^4$ and $S^2 \times S^2$
for the cases $m = 0$ and $m = m_c$, respectively. The case $m =
0$ leads to the creation of a universe without a black hole and
the case $m = m_c$ leads to the creation of a universe with a
pair of maximal black holes [6]. For the general case, one can
make a
constrained instanton as follows. The constrained
instanton is the seed for the quantum creation of a
Schwarzschild-de
Sitter black hole pair, or a sub-maximal black hole pair [3], and
$r_2$
and $r_3$ become the black hole and cosmological horizons for the
holes created. 

One can have two cuts at $\tau = \pm \Delta \tau /2$ between the
two horizons. Then the $f_2$-fold cover around the black hole
horizon $r = r_2$ turns the $(\tau - r)$ plane into a cone with a
deficit angle $2\pi (1 - f_2)$ there. In a similar way one
can have an $f_3$-fold cover at the cosmological horizon. In
order to form a fairly symmetric Euclidean manifold, one can glue
these two cuts under the condition 
\begin{equation}
f_2 \beta_2 + f_3 \beta_3 = 0,
\end{equation}
where $\beta_i = 2\pi \kappa_i^{-1}$  are the periods of $\tau$
that avoid conical
singularities in compacting the Euclidean spacetime
at these two horizons, respectively. The absolute values of their
reciprocals are the
Hawking temperature and the Gibbons-Hawking temperature.
If $f_2$ or $f_3$ is different from $1$ (at least one should
be,
since the two periods are different for the sub-maximal black
holes), then the cone at the black hole or cosmological horizon
will have an extra contribution to the action of the instanton.
After the transition to Lorentzian spacetime, the conical
singularities will only affect the real part of the phase of the
wave function, i.e. the probability of the black hole creation.

The extra contribution due to the conical singularities can be
considered as the degenerate form of the surface term in the
action (2) and can be written as follows:
\begin{equation}
I_{i,deficit} = - \frac{1}{8 \pi}\cdot 4\pi r_i^2\cdot 2\pi
(1 - f_i).\;\;\; (i = 2, 3)
\end{equation}

The volume term of the action for the instanton can be calculated
\begin{equation}
I_{vol} = -\frac{\Lambda}{6} (r^3_3 - r^3_2) f_2 \beta_2.
\end{equation}

Using eqs. (11) - (14), one obtains the total action
\begin{equation}
I = - \pi (r^2_2 + r^2_3).
\end{equation}
This is one quarter of the negative of the sum of the two horizon
areas. One quarter of the sum is the total entropy of the
universe. 
 
It is remarkable to note that the action is independent of the
choice of $f_2$ or $f_3$. Our manifold satisfies the Einstein
equation everywhere except for the two horizons at the equator.
The equator is two joint sections $\tau = consts. $ passing 
these horizons. It divides the instanton into two halves. The
Lorentzian metric of the black hole pair created can be obtained
through an analytic continuation of the time coordinate from
an imaginary to real value at the equator.  
We can impose the  restriction that the 3-geometry
characterized
by the parameter $m$ is given at the equator, i.e. the transition
surface.
The parameter $f_2$ or $f_3$ is the only degree of freedom left,
since the field equation holds elsewhere. Thus, in order to check
whether we get a stationary action solution for the given
horizons, one only needs to see whether the above action is
stationary with respect to this parameter. Our result (15) shows
that our gravitational action has a stationary action and the
manifold is qualified as a constrained instanton. The exponential
of the negative of the action can be
used for the $WKB$ approximation to the probability.

Eq. (15) also implies that no matter which value of $f_2$ or
$f_3$ is
chosen,
the same black hole should be created with the same probability.
Of course, the most dramatic case  is the creation of a universe
from no volume, i.e. $f_2
=f_3 = 0$. 

From eq. (15) it follows that the relative probability of the
pair
creation of black holes in the de Sitter background is 
the exponential of the total entropy of the universe [3] [8].

One can study quantum no-boundary states of scalar and spinor
fields in
this model. It turns out that these fields are in thermal
equilibrium with the background. The
associated temperature is the reciprocal of the identification
time period as
expected, and it can take an arbitrary value [9].

Now, let us discuss the case of $\Lambda < 0$. One is interested
in the probability of pair creation
of Schwarzschild-anti-de Sitter black holes. The universe is
open.
Hence, our key point is to find a
complex solution which has both the universe as its Lorentzian
section and a compact section as the seed for the creation, i.e.
the constrained instanton. The real part of its
action will determine the creation probability.

The metric of the constrained instanton takes the same form as
eq. (5). However, two zeros of $\Delta$ become complex
conjugates.
One can define
\begin{equation}
\gamma \equiv \frac{1}{3} \mbox{arcsinh} (3m
|\Lambda |^{1/2})),
\end{equation}
and then one has
\[
r_2 = 2\sqrt{\frac{1}{|\Lambda|}} \sinh \gamma,
\]
\begin{equation}
r_3 = \bar{r}_0 = \sqrt{\frac{1}{|\Lambda |}} (  - \sinh \gamma -
i \sqrt{3} \cosh \gamma ).
\end{equation}

One can build a complex constrained instanton using the section
connecting $r_0$ and $r_3$. Since
$r_0$ and $r_3$ are complex conjugates, the real part of $r$ on
the section is constant, and the
range of the imaginary part runs between $ \pm
i\sqrt{\frac{3}{|\Lambda |}}  \cosh \gamma $. The surface
gravities $\kappa_0$ and $\kappa_3$ are
complex conjugates too.
Following the procedure of constructing the constrained
gravitational instanton for the case
$\Lambda > 0$, we can use complex folding parameters $f_0$ and
$f_3$ to
cut, fold and glue the
complex manifold with
\begin{equation}
f_0 \beta_0  + f_3 \beta_3 = 0.
\end{equation}

As expected, the action is independent of the parameter $f_0$ or
$f_3$ and
\begin{equation}
I = - \pi ( r_0^2 + r_3^2) = \pi \left (-\frac{6}{\Lambda} +
r^2_2 \right ).
\end{equation}

The action is independent from the choice of the time
identification period.
One can always choose the arbitrary time identification period
to be imaginary, and the Lorentzian section in which we are
living is associated with the real time. A special choice of the
imaginary time period will regularize the conical singularity of
the Euclidean section at the black hole horizon $r_2$. However,
we
do not have to do so, since constrained instantons are allowed
in quantum cosmology.

One can obtain the Lorentzian metric from an analytic
continuation
of the time coordinate from an imaginary to real value at the
equator of the
instanton. The equator is two joint $\tau = consts.$ sections
passing through
these horizons. The 3-geometry of the equator can be considered
as the restriction
imposed for the constrained instanton. Again, the independence of
(19) from
the time identification period shows that the manifold is
qualified as a
constrained instanton. 

Therefore, the relative probability of the pair creation of
Schwarzschild-anti-de Sitter black holes, at
the $WKB$ level, is the exponential of the negative of one
quarter of the black hole horizon area,  in contrast to
the case of pair creation of black holes in the de Sitter space
background. One quarter of the black hole horizon area is known
to be the entropy in the Schwarzschild-anti-de Sitter universe
[10].

One may wonder why we choose horizons $r_0$ and $r_3$ to
construct the instanton. One
can also consider those constructions involving $r_2$ as the
instantons. However, the real part
of the action for our choice is always greater
than that of the other choices for the given configuration, and
the wave function or the probability is determined by the
classical orbit
with the greatest real part of the action [1]. When we dealt with
the Schwarzschild-de Sitter case, the choice of the instanton
with $r_2$ and $r_3$ had the greatest action accidentally, but we
did not appreciate this earlier.

\vspace*{0.3in}

\large \bf III. Reissner-Nordstr$\rm\bf\ddot{o}$m
\vspace*{0.15in}

\rm 

\normalsize
 
Now, let us include the Maxwell field into the model. The
Reissner-Nordstr$\rm\ddot{o}$m-de Sitter spacetime, with
mass
parameter $m$, charge $Q$ and a positive  cosmological
constant $\Lambda$, is the only spherically
symmetric electrovac solution to the Einstein and Maxwell
equations. 
Its Euclidean metric can be written as  
\begin{equation} 
ds^2 = \Delta d\tau^2  
+ \Delta^{-1}dr^2 
+ r^2 (d\theta^2 + \sin^2 \theta d\phi^2),
\end{equation} 
where $\Delta$ is
\begin{equation}
\Delta = \left (1- \frac{2m}{r} +\frac{Q^2}{r^2} - \frac{\Lambda
r^2}{3} \right).
\end{equation} 
 
For convenience one can make a factorization  
\begin{equation} 
\Delta \equiv - \frac{\Lambda}{3r^2} (r - r_0)(r - r_1)(r -
r_2)(r -r_3), 
\end{equation} 
where $r_0, r_1, r_2, r_3$ are roots of $\Delta$ in ascending
order.

The gauge field is
\begin{equation}
 F= - \frac{iQ}{r^2} d\tau \wedge dr
\end{equation}
for an electrically charged solution, and
\begin{equation}
F= Q\sin \theta d \theta \wedge d \phi
\end{equation}
for a magnetically charged solution. We shall not consider dyonic
solutions. 

The roots satisfy the following relations
\begin{equation}
\sum_i r_i = 0,
\end{equation}
\begin{equation}
\sum_{i >j} r_i r_j = -\frac{3}{\Lambda},
\end{equation}
\begin{equation}
\sum_{i>j>k} r_i r_j r_k = -\frac{6m}{\Lambda}
\end{equation}
and
\begin{equation}
\prod_i r_i = - \frac{3Q^2}{\Lambda}.
\end{equation}

If all the roots $r_0, r_1, r_2, r_3$ are real, then $r_0$ is
negative, $r_2$ and $r_3$ are identified as the inner and outer
black hole horizons, and $r_3$ is the cosmological horizon.

The  surface gravity $\kappa_i$ of $r_i$ is [7] 
\begin{equation}
\kappa_i = \frac{\Lambda}{6r_i^2}\prod_{j = 0,1,2,3, \;\; (j
\neq i)}(r_i - r_j).
\end{equation}

One can make a real constrained gravitational instanton by
gluing two sections of constant values of imaginary time $\tau$
between the two complex horizons $r_1$ and $r_2$.  Then 
the $f_i$-fold $(i = 1, 2)$ cover turns 
the $(\tau - r)$ plane into a cone with a deficit angle 
$2\pi (1-f_i)$ at the horizons.  Both $f_1$ and $f_2$ can take
any pair of complex numbers with the relation  
\begin{equation} 
f_1 \beta_1 + f_2 \beta_2 = 0, 
\end{equation} 
where $\beta_i = 2\pi \kappa^{-1}_i$.
If $f_1$ or $f_2$ is different from $ 1$, then the cone at the
complex horizon will have an extra 
contribution to the action of the instanton. One can choose $f_1$
to make the
time identification period to be imaginary. The Lorentzian metric
for the black hole pair created can be obtained by analytic
continuation of the imaginary time to real time at the equator of
the constrained instanton. The equator is two joint sections
$\tau = consts. $ passing
the two horizons. It divides the instanton into two halves, and
has topology
$S^2 \times S^1$. After the transition
to Lorentzian spacetime, the conical singularities will affect 
the real part of the phase of the wave function, i.e. the
probability of the creation of the black holes.

The contributions to the action due to the conical singularities
at
the horizons are 
\begin{equation} 
I_{i,deficit} = - \frac{1}{8 \pi}\cdot 4\pi r_i^2\cdot 2\pi
(1 - f_i). \;\; (i = 1,2)
\end{equation} 
They are degenerate forms of the surface terms.

The action due to the volume is 
\begin{equation}
I_v = -\frac{f_1 \beta_1 \Lambda}{6} (r^3_2 - r^3_1) \pm
\frac{f_1 \beta_1 Q^2}{2}(r^{-1}_1 - r^{-1}_2),
\end{equation}
where $+$ is for the magnetic case and $-$ is for the electric
case. 

In the magnetic case, the boundary date is
$h_{ij}$ and $A_i$. The vector potential, in turn, determines the
magnetic charge, since it can be obtained by the magnetic flux,
or the integral of the gauge field $F$ over the $S^2$ factor. It
is more convenient to choose a gauge potential
\begin{equation}
A = Q(1- \cos \theta) d\phi
\end{equation} 
to evaluate the flux.

In the electric case, the boundary data is $h_{ij}$ and the
momentum $\omega$ [3][11][12], which is canonically conjugate to
the electric charge and defined by
\begin{equation}
\omega = \int A,
\end{equation}
where the integral is around the $S^1$ direction. The most
convenient choice of the gauge potential for the calculation is
\begin{equation}
A = -\frac{iQ}{r^2}\tau dr.
\end{equation}

The wave function for the equator is the exponential of half the
negative of the
action. For the  magnetic cases, one obtains the wave
function $\Psi (h_{ij})$ and $\Psi (Q,h_{ij})$.
For the electric case, what one obtains this way is
$\Psi (\omega, h_{ij})$ instead of $\Psi (Q, h_{ij})$. One can
get the wave function $\Psi (Q, h_{ij})$ for
 a given electric charge through the Fourier
transformation 
\begin{equation}
\Psi (Q, h_{ij}) = \frac{1}{2\pi} \int^{\infty}_{-\infty} d
\omega e^{i\omega Q} \Psi
(\omega, h_{ij}).
\end{equation}
This Fourier transformation is equivalent to a multiplication of
an extra factor
\begin{equation}
\exp \left (\frac{- f_1\beta_1Q^2  ( r_1^{-1} -
r_2^{-1})}{2}\right )
\end{equation}
to the wave function.
This makes the probabilities for magnetic and electric cases
equal, and thus recovers the duality between the magnetic and
electric black holes [3][11][12].

Therefore, the effective action for both the magnetic and
electric case is
\begin{equation}
I = - \pi (r^2_1 + r^2_2).
\end{equation}

The effective action is independent from the choice of the time
identification period. By the same argument, the manifold
constructed is qualified as a constrained instanton.

The relative probability of the universe creation is the
exponential of the
negative of the seed instanton action. In our case,  this is the
exponential of the sum of the outer and inner black hole horizon
areas.

It is noted that if one has not used the Fourier transformation
(36), then one cannot obtain the constrained instanton and the
whole calculation becomes meaningless.

The reason that we use the inner and outer black hole horizons to
construct the constrained instanton is
that the instanton has the largest action in comparison with
other
options, as can be simply proven. For the same configuration the
orbit with the largest real part of the
action determines the wave function and probability [1].

Now let us discuss the pair creation in the anti-de Sitter space
background. 
The spherically symmetric electrovac model with a negative
cosmological constant is also described by eqs. (20)-(24). For
the
general case, one has two real horizons $r_2, r_3$, which are
identified as the inner and outer black hole horizons. The other
two 
horizons $r_0, r_1$ are complex conjugates. One then uses these
two complex horizons to construct the complex constrained
instanton as in the Schwarzschild-anti-de Sitter case.  The
action  is
\begin{equation}
I = -\pi (r^2_0 + r_1^2) = \pi \left ( - \frac{6}{\Lambda} +
r^2_2 + r_3^2 \right ),
\end{equation}
where we use the fact that the sum of all horizon areas is equal
to
$24\pi \Lambda^{-1}$, which is implied by eqs. (25)(26).

The relative creation probability is the exponential of the
negative of the seed instanton action. In our case, this becomes
the exponential of the negative of the sum of the outer and inner
black hole horizon areas.

This choice of the horizons for the construction is again
justified by the fact that, the action of the instanton
considered is largest for the configuration of the wave function
[1]. This point is important. For
example, if, instead we use $r_2$ and $r_3$ for constructing the
instanton, then the creation probability of a universe without a
black hole would be smaller than that with a pair of black holes.
This is physically absurd.

The pair creation of Reissner-Nordstr$\rm\ddot{o}$m-de
Sitter black holes from regular instantons was studied in
[11][12][13][14]. These become the special cases of the
discussion presented in [3]. The pair creation of compactified
Reissner-Nordstr$\rm\ddot{o}$m-anti-de
Sitter black holes was discussed in [15][16]. However, to avoid
the difficulty associated with open creation, a domain wall was
introduced to compactify the non-compact geometry.

\vspace*{0.3in}

\large \bf IV. Kerr
\vspace*{0.15in}

\rm 

\normalsize

Now let us discuss the creation of a rotating black hole
in the (anti-)de Sitter space background. The Lorentzian metric
of the black hole spacetime is [7]
\begin{equation}
ds^2 = \rho^2(\Delta^{-1}_r dr^2 + \Delta^{-1}_\theta d\theta^2)
+ \rho^{-2}
 \Xi^{-2}
\Delta_{\theta} \sin^2 \theta (adt - (r^2 + a^2) d\phi)^2 -
\rho^{-2} \Xi^{-2}\Delta_r  (dt - a \sin^2 \theta d \phi)^2,
\end{equation}
where
\begin{equation}
\rho^2 = r^2 + a^2 \cos^2 \theta,
\end{equation}
\begin{equation}
\Delta_r = (r^2 + a^2)(1 - \Lambda r^2 3^{-1}) - 2mr + Q^2 + P^2,
\end{equation}
\begin{equation}
\Delta_{\theta} = 1 + \Lambda a^2 3^{-1} \cos^2 \theta,
\end{equation}
\begin{equation}
\Xi = 1 + \Lambda a^2 3^{-1}
\end{equation}
and $m, a, Q$ and $P$ are constants, $m$ and $ma$ representing
mass and  angular momentum. $Q$ and $P$ are electric and
magnetic charges.

One can factorize $\Delta_r$ as follows
\begin{equation}
\Delta_r = -\frac{\Lambda}{3} (r - r_0)(r - r_1)(r - r_2)(r -
r_3).
\end{equation}

The roots $r_i$ satisfy the following relations:
\begin{equation}
\sum_i r_i = 0,
\end{equation}
\begin{equation}
\sum_{i>j} r_i r_j = - \frac{3}{\Lambda} + a^2,
\end{equation}
\begin{equation}
\sum_{i>j>k} r_ir_jr_k = - \frac{6m}{\Lambda},
\end{equation}
\begin{equation}
\prod_i r_i = - \frac{3(a^2 + Q^2 + P^2)}{\Lambda}.
\end{equation}

We shall concentrate on the neutral case with $Q
= P = 0$ first. The charged case with nonzero
electric or magnetic charge will be discussed later.

For the general closed case with a positive cosmological
constant, one root, say $r_0$, is negative, one can identify the
other three positive roots $r_1, r_2, r_3$ as the inner black
hole, outer black hole and cosmological  horizons, respectively.

The probability of the Kerr-de Sitter black hole pair 
creation, at the $WKB$ level, is the
exponential of the negative of the action of its constrained
gravitational instanton.  
In order to form a constrained gravitational instanton, one can
do the similar cutting, folding and covering between the inner
and outer black hole horizons
with folding parameters $f_1$ and $f_2$
satisfying relation (30) as in the nonrotating case. The reason
to consider the construction with horizons $r_1$ and $r_2$ as the
seed instanton is the same: It has the largest action for the
same configuration of the wave function comparing with other
choices. The
Lorentzian metric for the black hole pair created is obtained
through analytic continuation in the same way as for the
nonrotating case.
The equator where the quantum transition will occur has topology
$S^2 \times S^1$.
The restrictions can be similarly imposed at the equator for the
constrained instanton.

The horizon areas are
\begin{equation}
A_i = 4\pi (r^2_i + a^2)\Xi^{-1}.
\end{equation}

The surface gravities of the horizons are
\begin{equation}
\kappa_i =\frac{\Lambda \prod_{j\; (j \neq i)}(r_i - r_j)}{6\Xi
(r^2_i + a^2)}.
\end{equation}

The actions due to the horizons are
\begin{equation}
I_{i, horizon} = - \frac{\pi (r^2_i + a^2)(1 -
f_i)}{\Xi}.\;\; (i = 1,2)
\end{equation}
 
The action due to the volume is
\begin{equation}
I_v = - \frac{f_1\beta_1 \Lambda}{6\Xi^2} (r^3_2 - r^3_1 +
a^2(r_2 - r_1)),
\end{equation}
where we define $\beta_i= 2\pi \kappa_i^{-1}$.

If one naively takes the exponential of the negative of half the
total action, then the exponential is not identified as the wave
function at the 
creation moment of the black hole pair. The
physical reason is that what one can observe is only the angular
differentiation, or the relative rotation of the two horizons.
This situation is similar to the case of a Kerr black hole pair
in the
asymptotically flat background. There one can only measure the
rotation of
the black hole horizon from the spatial infinity. To find the
wave function for the given mass and angular momentum one has to
make the Fourier transformation [3]
\begin{equation}
\Psi(m, a, h_{ij}) = \frac{1}{2 \pi}\int^{\infty}_{-\infty}
d\delta e^{i\delta
J \Xi^{-2}} \Psi(m, \delta, h_{ij}),
\end{equation}
where $\delta$ is the relative rotation angle for the half time
period
$f_1\beta_1/2$, which is canonically conjugate to the angular
momentum $J = ma$; and the factor $\Xi^{-2}$ is due to the
time rescaling.
The angle difference $\delta$ can be evaluated
\begin{equation}
\delta = \int_0^{f_1\beta_1/2} d\tau (\Omega_1 - \Omega_2),
\end{equation}
where the angular velocities at the horizons are
\begin{equation}
\Omega_i = \frac{a}{r^2_i + a^2}.
\end{equation}

The Fourier transformation is equivalent to adding an extra term
into the action for the constrained instanton, and then the total
action becomes
\begin{equation}
I = - \pi(r^2_1 + a^2)\Xi^{-1} - \pi(r^2_2 + a^2)\Xi^{-1}.
\end{equation}

It is crucial to note that the action is independent of
the identification time period $\beta$, $f_1 \beta_1$ for our
case, and therefore, the manifold obtained is qualified
as a constrained instanton. Again, if one does not adapt the
Fourier transformation between the angular momentum and the
relative rotation angle, then one cannot obtain the constrained
instanton. 

Therefore, the relative probability of the Kerr black hole pair
creation is
\begin{equation}
P_k \approx \exp  (\pi(r^2_1 + a^2)\Xi^{-1} + \pi(r^2_2 +
a^2)\Xi^{-1}).
\end{equation}
It is the exponential of one quarter of the sum of the outer
and inner black hole horizon areas.

For the general open case with a negative cosmological constant,
there exist at least two complex conjugate roots, say $r_0, r_1$.
We assume the other two roots $r_2, r_3$ to be real and identify
them as the inner and outer black hole horizons. By the
same reason, in order to obtain an instanton with the greatest
action for a given configuration, one has to choose complex
horizons $r_0, r_1$ to make a  constrained instanton.

After the Fourier transformation associated with the rotation,
one can show that the construction is indeed the instanton
required. It is the seed for the pair creation of Kerr-anti-de
Sitter black holes. The action is
\begin{equation}
I = - \pi(r^2_0 + a^2)\Xi^{-1} - \pi(r^2_1 + a^2)\Xi^{-1}
=\pi \left ( -\frac{6}{\Lambda}+ (r^2_2 + a^2)\Xi^{-1} +(r^2_3 +
a^2)\Xi^{-1} \right ),
\end{equation}
where we use the fact that the sum of all horizon areas is $24\pi
\Lambda^{-1}$, which can be derived from eqs. (46)(47).

Therefore, the  relative probability is
\begin{equation}
P_k \approx \exp  -(\pi(r^2_2 + a^2)\Xi^{-1} + \pi(r^2_3 +
a^2)\Xi^{-1}).
\end{equation}
It is the exponential of the negative of one quarter of the sum
of the inner and outer black hole horizon areas.

\vspace*{0.3in}

\large \bf V. Newman
\vspace*{0.15in}

\rm 

\normalsize

Now, let us turn to the charged and rotating black hole case. The
vector potential can be written as 
\begin{equation}
A =\frac{ Qr(dt - a\sin^2\theta d\phi) + P \cos \theta (a dt -
(r^2 + a^2) d\phi)}{\rho^2}.
\end{equation}

We shall not consider the dyonic case in the following.

One can closely follow the neutral rotating case for calculating
the action of the corresponding constrained gravitational
instanton. The only difference is to add one more term due to the
electromagnetic field to the action of volume. For the magnetic
case, it is
\begin{equation}
\frac{f_1\beta_1 P^2}{2\Xi^2} \left ( \frac{r_i}{r^2_i+ a^2} -
\frac{r_j}{r^2_j + a^2} \right )
\end{equation}
and for the electric case, it is
\begin{equation}
-\frac{f_1\beta_1 Q^2}{2\Xi^2} \left ( \frac{r_i}{r^2_i+ a^2} -
\frac{r_j}{r^2_j + a^2} \right ),
\end{equation}
where $(i,j)$ is $(1,2)$ and $(0,1)$ for the closed and open
cases, respectively.

In the magnetic case the vector potential determines the magnetic
charge, which is the integral  over the $S^2$ factor.
Putting
all these contributions together one can find
\begin{equation}
I = - \pi(r^2_i + a^2)\Xi^{-1} - \pi(r^2_j + a^2)\Xi^{-1},
\end{equation}
and the relative probabilities of the closed and open pair
creations of magnetically
charged black holes is written in the same form as eqs. (58)(60).

In the electric case, one can only fix the integral
\begin{equation}
\omega = \int A,
\end{equation}
where the integral is around the $S^1$ direction.

So, what one obtains in this way is $\Psi(\omega, a, h_{ij})$. In
order to get the wave function $\Psi(Q, a, h_{ij})$ for a given
electric charge, we have to repeat the procedure like the
Reissner-Nordstr$\rm\ddot{o}$m case. The Fourier transformation
is equivalent to adding one more term to the action
\begin{equation} 
\frac{f_1\beta_1 Q^2}{\Xi^2}\left ( \frac{r_i}{r^2_i+ a^2} -
\frac{r_j}{r^2_j + a^2} \right ).
\end{equation} 
Then we obtain the same probability formulas for the electrically
charged rotating black hole pair creation as for the
magnetic one. The duality between the magnetic and electric cases
is recovered.

The pair creation of Kerr-Newman-de
Sitter black holes from a regular instanton was also studied 
[17], as a special case of the general discussion [3].

\vspace*{0.3in}

\large \bf VI. Thermodynamics
\vspace*{0.15in}

\rm 

\normalsize

In the constrained instanton approach to quantum cosmology
associated with black hole pair creation, the fact that the
action is independent from the imaginary time period
$\beta$ is crucial.

In gravitational thermodynamics, the partition function $Z$ is
identified with the path integral over all metrics $g$ and matter
fields $\psi$ on a manifold [18], 
\begin{equation}
Z = \int d[g] d[\psi] \exp - I(g, \psi ).
\end{equation}

The $WKB$ approximation to the path integral is equivalent
to the contribution of the
background, excluding the fluctuations. The background is the
stationary action orbit, or the constrained instanton.  At
this level, one has
\begin{equation}
Z = \exp -I.
\end{equation}

For the regular compact instanton case, 
there are no externally imposed quantities and corresponding
chemical potentials. Therefore, the partition function simply
counts the total number of the states, and each state is equally
probable with the probability $p_n = Z^{-1}$. Thus, the entropy
is $S = -p_n \log p_n = \log Z$. Therefore, at the $WKB$ level,
the entropy is the negative of the action of the instanton [19].

For the regular noncompact instanton case, like the 
Kerr-Newman-anti-de Sitter family, the system is constrained by
three quantities, namely mass or energy $m$, electric charge
$Q$
and angular momentum $J$. There are two corresponding chemical
potentials, the electrostatic potential $\Phi$ at the outer black
hole horizon and
angular velocity $\Omega$. Then one
has to use the grand partition function $Z$ in grand canonical
ensembles for the thermodynamics study [18],
\begin{equation}
Z = \mbox{Tr} \exp (-\beta m + \beta \Omega J + \beta \Phi Q)
=\int
d[g] d[\psi] \exp - I(g, \psi ).
\end{equation}
Here, the path integral is over all fields whose value at the
point $(\tau- \beta, r, \theta, \phi + i\beta \Omega)$ is $\exp
(Q\beta, \phi)$ times the value at $(\tau, r, \theta, \phi )$
[20].

Strictly speaking, the quantity $\beta$ of the term $\beta m$ is
the  difference
of the Euclidean time lapse at infinity, which is $\beta$, and
that at the outer black hole horizon, which is zero.  The
quantity $\beta \Omega J$ is the difference of the rotation
angles measured at the horizon and at infinity during the
Euclidean time lapses. The quantity $\beta \Phi$ is the
difference of the electrostatic potentials at the horizon and at
infinity.

For the constrained instanton case, the imposed quantities are
given by the constraints for the instanton. The partition
function (69) remains valid, but the relevant quantities should
be
interpreted as the differences between that at the two horizons
used for constructing the instanton, instead of that at the outer
black hole horizon and at infinity. Since the Euclidean time
lapses
at two horizons are zero, the term $\beta m$ disappears in the
exponent of (69). 

The dominant contribution to the path integral (69) is due to a
stationary action orbit. However, there does not exists such an
orbit satisfying the jump condition for nonzero
$Q$ or $J$. It is noted that $a$ is real. Instead, the $WKB$
approximation of the path integral is the exponential of the
negative of the effective action of the constrained instanton
constructed earlier. The instanton is obtained from the cutting,
folding, gluing of the complex solution. If needed, the two
Fourier transformations are introduced. This is an alternative
justification of the two
Fourier transformations.

Even for the compact instanton case, if the instanton is
electrically charged, or rotating, or both, then the naively
evaluated partition
function  without imposed quantities as in (67),
does not correspond to what we want. One has to impose the
quantities $Q$, or $J$, or both, and use (69) instead.

The entropy $S$ can be obtained by
\begin{equation}
S = - \frac{\beta \partial}{\partial \beta} \ln Z + \ln Z = - I,
\end{equation}
where the action $I$ is the effective one, of course.

Thus, the condition that $I$ is independent from $\beta$ implies
that the ``entropy'' is the negative of the action. 

For the closed creation of the chargeless and nonrotating black
holes, the ``entropy'' is the true entropy of the universe. For
the
closed creation of the charged or (and) rotating black holes, the
``entropy'' is one quarter of the sum of the inner and outer
black hole horizon areas. For compact regular instantons, the
fact that the entropy is equal to the negative of the action is
shown using different arguments in [19].

It is noted from eqs. (25)(26)(46) and (47) that, for all members
of the Kerr-Newman-(anti-)de Sitter family, the sum of all
horizon areas is equal to $24\pi\Lambda^{-1}$. This fact seems
coincidental, but it has a deep physical
significance.

For the open creation of the chargeless and nonrotating black
holes, the ``entropy'' is associated with the complex horizons.
It becomes the negative of the true entropy of the universe, up
to a
constant $6\pi \Lambda^{-1}$. Equivalently, the action is the
entropy up to the same constant. This constant is ignored in the
background subtraction approach anyway. For the open creation of
the charged or (and) rotating black holes, the ``entropy''
becomes
one quarter of the negative of the sum of the inner and outer
black hole horizon areas up to the constant.

Using a standard technique designed for spaces
with spatially noncompact geometries [18], the action of the
Kerr-Newman-anti-de Sitter space is evaluated as follows:  The
physical action is defined by the difference between the
action of the space under study and that of a
reference background. The background can be a static solution to
the field equation. From gravitational thermodynamics, one can
derive the entropy from the action.

In quantum gravity the quantum state can be represented by a
matrix density. Apparently, the state associated with our
constrained instanton is  an eigenstate of
the entropy operator, instead of the 
temperature operator, as previously thought.

\vspace*{0.3in}

\large \bf VII. Discussion
\vspace*{0.15in}

\rm 

\normalsize

The Hawking temperature is defined as the reciprocal of the
absolute value of the time identification period required to make
the Euclidean manifold regular at the horizon. In the background
subtraction approach for an open universe, if one lifts the
regularity condition at the horizon, or
lets the period take an arbitrary  value, then one finds
that the action does depend on the
period and becomes meaningless. However, if we calculate the
action using our complex
constrained instanton, then the action is independent of the
complex period $ \beta$. It is noted that the values of the
action are different for these two methods. The beautiful aspect
of our approach is that,
even in the absence of a general no-boundary proposal for open
universes, we can treat the creation of the
closed and the open universes in the same way.

Our treatment of quantum creation of the
Kerr-Newman-anti-de Sitter space using the constrained
instanton can
be thought of as a prototype of quantum gravity for an open
system without appealing to the background subtraction approach.

The Kerr-Newman black hole case can be thought of as the limit
of our case as we let $\Lambda$ approach  $0$ from below. The
constrained
instanton approach naturally explains the  fact
that the action and entropy of a Schwarzschild space are 
equal.

In the constrained instanton approach, we use the
effective action after introducing the
two Fourier transformations. This is equivalent to the
requirements for the proper evaluation of (69) in  grand
canonical ensembles. Since the effective
action is
independent of $\beta$, then the ``entropy'' $S$ becomes the
negative of the effective action $I$. However, one has to clarify
the meaning of the ``entropy.''

When we construct the constrained instanton, we always select two
horizons such that the action takes its largest value. Only for
the case of Schwarzschild-de Sitter black hole pair creation,
does the quantum
transition occur at a true spacelike 3-surface. For the rest
of the cases, the transitions seem quite counter-intuitive.
People have
spent a lot of effort developing quantum gravity for spacetimes
with a
$U(1)$ isometry. The Killing time associated with the isometry is
imaginarized to obtain the Euclidean orbits. This procedure is
justified by the isometry. However, for general spacetimes, there
does not exist a prestigious time coordinate. From the principle
of general covariance, one has to work in the domain of all
complex
coordinates. This is the reason that one has to live
with the quantum transition through any kind of 3-surface.

In summary, for the chargeless and
nonrotating case, the closed (open) creation probability is the
exponential of the (minus) entropy of the universe, and for the
other cases (charged, rotating, or both), the creation
probability is the
exponential of (minus) one quarter of the sum of the inner and
outer black hole horizon areas.

Due to the black hole No-Hair Theorem, the problem of
black hole pair creation in both the de Sitter and the anti-de
Sitter space backgrounds has been completely resolved.

It can be shown that  the probability of  the universe creation
without a black
hole is greater than that with a pair of black holes in both
the de Sitter and the anti-de Sitter backgrounds.

\vspace*{0.3in}

\bf References:

\vspace*{0.1in}
\rm

1. J.B. Hartle and S.W. Hawking, \it Phys. Rev. \rm \bf D\rm
\underline{28}, 2960 (1983).

2. S.W. Hawking and N. Turok, \it Phys. Lett. \rm \bf B\rm
\underline{425}, 25 (1998), hep-th/9802030.

3. Z.C. Wu, \it Int. J. Mod. Phys. \rm \bf D\rm\underline{6}, 199
(1997), gr-qc/9801020.

4. Z.C. Wu,  \it Phys. Rev. \rm \bf D\rm
\underline{31}, 3079 (1985).

5. Z.C. Wu, Beijing preprint, hep-th/9803121.

6. R. Bousso and S.W. Hawking, \it Phys. Rev. \rm \bf D\rm
\underline{52}, 5659 (1995), gr-qc/9506047.

7. G.W. Gibbons and S.W. Hawking, \it Phys. Rev. \bf D\rm
\underline{15}, 2738 (1977).

8. R. Bousso and S.W. Hawking, hep-th/9807148.
 
9. Z.C. Wu, Beijing preprint, gr-qc/9712066.

10. S.W. Hawking and D.N. Page, \it Commun. Math. Phys. \rm
\underline{87}, 577 (1983).

11. F. Mellor and I. Moss, \it Phys. Lett. \bf B\rm 
\underline{222}, 361 (1989).

12. I.J. Romans, \it Nucl. Phys. \bf B\rm 
\underline{383}, 395 (1992).

13. R.B. Mann and S.F. Ross, \it Phys. Rev. \bf D\rm
\underline{52}, 2254 (1995). 
 
14. S.W. Hawking and S.F. Ross,  \it Phys. Rev. \bf D\rm 
\underline{52}, 5865 (1995).

15. R.R. Caldwell,  A. Chamblin and G.W. Gibbons, \it Phys. Rev.
\bf D\rm \underline{53}, 7103 (1996).

16. R.B. Mann, \it Class. Quantum Grav. \rm 
\underline{14}, L109 (1997).

17. I.S. Booth and R.B. Mann, gr-qc/9806015.

18. S.W. Hawking,  in \it General Relativity: An Einstein
Centenary Survey, \rm eds. S.W. Hawking and W. Israel, (Cambridge
University Press, 1979).

19. G.W. Gibbons and S.W. Hawking, \it Commun. Math. Phys. \rm
\underline{66}, 291 (1979).

20. G.W. Gibbons and S.W. Hawking, \it Phys. Rev. \bf D\rm
\underline{15}, 2752 (1977).

\end{document}